# SOCRATES

## A System For Scalable Graph Analytics

*C. Savkli, R.Carr, M. Chapman, B. Chee, D. Minch*
Johns Hopkins University, Applied Physics Laboratory
Mail Stop MP6-N312
Laurel, MD 20723-6099 USA
cetin.savkli@jhuapl.edu

*Abstract*— **A distributed semantic graph processing system that provides locality control, indexing, graph query, and parallel processing capabilities is presented.**

*Keywords—graph, semantic,*

## I. INTRODUCTION

Graphs provide a flexible data structure that facilitates fusion of disparate data sets. Popularity of graphs has shown a steady growth with the development of internet, cyber, and social networks. While graphs provide a flexible data structure, processing of large graphs remain a challenging problem.

Successful implementation of graph analytics revolves around several key considerations: rapid data ingest and retrieval, scalable storage, and parallel processing. In this paper we present a graph analytics platform that is particularly focused on facilitating large scale analysis of semantic graphs.

Recently, NoSQL systems such as Hadoop have become popular for storing Big Data, however these systems face several fundamental challenges that make analyzing that data difficult:

- Lack of secondary indexing, which leads to poor performance of attribute queries (e.g. "What flights have we seen moving faster than 500mph?").
- Lack of locality control, which can lead to unnecessary movement of data.
- Lack of schema, which makes database maintenance challenging.

More traditional relational databases (a.k.a. RDBMS) do not have these problems, but face their own set of challenges when dealing with Big Data:

- Table structures not flexible enough to support new kinds of data easily.
- Poor parallelization & scalability.

SOCRATES provides a solution that combines the key features of these two approaches and avoids all of the above problems. It features several advances in data management for parallel computing and scalable distributed storage:

- Graph API supports representation of different kinds of data using a simple, extensible database schema.
- Distributed storage & processing: Functionality to execute algorithms on each local machine and results are merged without moving data.
- Indexing: Every attribute is indexed for fast random access and analytics.
- Distributed management: Nothing in the cluster is centrally managed. This includes communication as well as locating graph elements.
- Locality control: Graph vertices can be placed on specific machines, a feature essential for minimizing data movement in graph analytics.
- Platform independence: Built using Java Standards and can run in heterogeneous hardware/software configurations.

## II. RELATED WORK

SOCRATES is designed for attribute rich data – having many labels on both nodes and edges. Systems such as Twitter's Cassovary and FlockDB, GPS, Pregel and Pegasus are fast at processing structural graphs, however, they do not leverage or provide any ability to store or use labels on edges or nodes besides edge weights [1][2][3][4][5]. Current graph benchmarking tools such as HPC Scalable Graph Analysis Benchmark (HPC-SGAB) generate tuples of data with the form <StartVertex, EndVertex, Weight> with no other attributes which plays well to the strengths of Cassovary or Pregel [6]. However, such benchmarks tend to ignore the functionality we specifically aim for within this work.

Other graph databases focus more on content or tend to model specific relationship types such as ontology. These databases include Jena, OpenLink Virtuoso, R2DF and other commercial offerings use Resource Description Framework (RDF), which was originally designed to represent metadata [6][7][8]. RDF expressions consist of triples with a subject, predicate and object that is stored and queried against. The predicate in each triple represents a relationship between a subject and object. Intuitively, a general set of RDF tuples can be considered a graph, however, formally RDF is not defined as a mathematical concept of a graph[9].

Neo4J, Titan, HypergraphDB and DEX offer similar capabilities to SOCRATES. SOCRATES, Neo4J and Titan for example make use of Blueprints API for interacting with graphs [10]. However, SOCRATES has extended this API to offer enhanced graph processing functionality that includes locality control, additional graph methods facilitating graph analytics, as well as a parallel processing capability.

SOCRATES is built upon a cluster of SQL databases, similar to FaceBook's TAO and Twitter's deprecated FlockDB [11][5]. FaceBook's social graph is currently stored in TAO, a data model and API specifically designed for social graphs. FaceBook's social graph is served via TAO which was tailor to fit its workload needs: frequent reads with fewer writes, most edge queries having empty results and node connectivity and data size have distributions with long tails. These long tails enable efficient cache implementations, for example if a large portion of reads are potentially in cache (people are interested in current events – time locality or alternatively something becomes "viral" and is viewed by many people). TAO enables relatively few queries and stores its main data objects and associations as key-value pairs. This is similar to how SOCRATES stores attributes utilizing many tables in which each attribute is stored as value and a node or edge's id as the key. This schema removes joins as a potential memory and performance bottleneck. TAO was developed with specific constraints in mind such as global scaling and time based social graphs, whereas SOCRATES is targeted at more general graph structures.

III. DESIGN

A. Data Management & Locality

SOCRATES uses the following conventions: Each graph vertex exists only on 1 machine and each graph edge can exist at most on 2 machines. Edges know the IDs of vertices they connect as well as the machine those nodes reside on. There is no central management of location information.

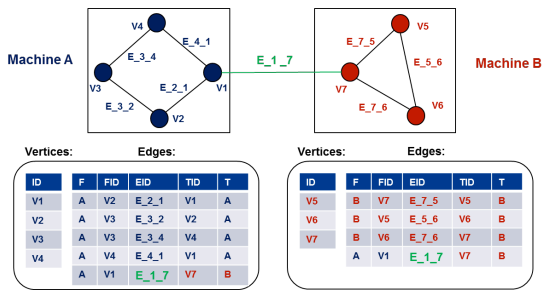

Fig. 1: Graph structure representation on the cluster.

Attributes of the graph are stored separately in 2 column tables where each attribute can be independently indexed and queried. This approach avoids the complexity typically associated with table structure changes in relational databases. Attributes of graph vertices are stored on the machine those vertices reside while edge attributes are stored on the machine where edges originate.

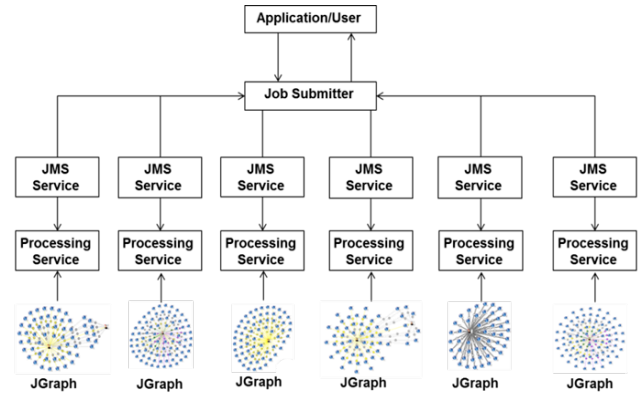

Fig. 2: The JGraph model of parallelism in Socrates. Users submit a processing job that is run in parallel on each node in the cluster, and given access to the subgraph stored on that node.

Access to data is provided by a Graph API where users can interact with the cluster using a simple graph interface. Graph API includes an implementation of Blueprints API. Besides the standard Blueprints methods SOCRATES graph API provides additional methods that take advantage of the architecture for efficient implementation of the analytics. One such method involves adding vertices of the graph to specific machines. The machine level access to graph allows a user to partition the graph to minimize edge crossings between machines. The following example, which is built using the Brightkite data set, illustrates the benefit of locality control:

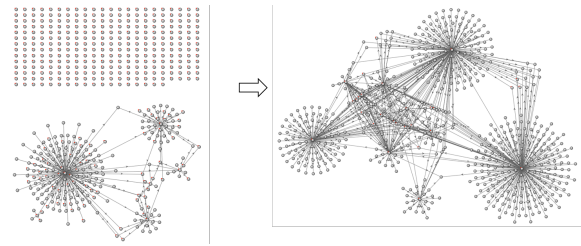

Fig. 3: Ability to partition graph over the cluster nodes minimizes movement of data.

The graph on the left represents 1 node of a 4 machine cluster where data was archived without locality control. In this case the probability that a neighbor resides on the same machine is ¼ which is reflected in the outcome where only about a quarter of vertices have their neighbors local. Example on the right shows what the partial graph on the same machine looks like when archived using SOCRATES. Ability to place graph nodes in particular machines can also be used to automatically partition the data when attributes can be hashed to generate a machine ID, such as partitioning of graph based on latitude-longitude attributes of vertices.

SOCRATES provides efficient implementation of parallelized Graph query for sub-graph matching. A particularly useful query involves finding joint neighbors of a pair of vertices. This query is a key operation for a variety of link discovery analysis and it is efficiently implemented without moving data irrespective of where vertices are located. Implementation of the joint neighbor finding relies on the data structure maintained on each machine to identify joint neighbors through a database query rather than iterating over neighbors of each vertex on the client. Every vertex on the graph knows the location and unique ID of vertices it is connected to without having to query another machine. An example of sub-graph query is provided in Figure 4. In this figure triangle shown on the left represent a graph query and matching result is illustrated on the right.

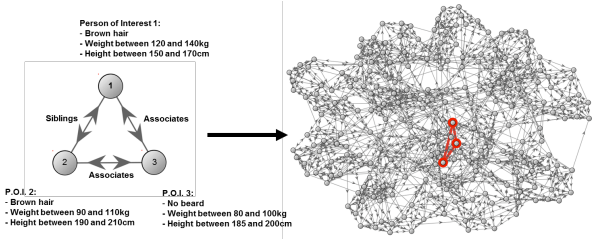

Fig. 4: A query example where a sub-graph with structure and attribute constraints is found in a larger graph.

## B. Parallel Processing

A key feature for handling large scale graphs is ability to process graph in parallel without having to move data. Primary challenge in parallel processing of graphs is associated with the fact that, for most nontrivial problems, analysis on each machine on the cluster requires access to data on other machines to be able to produce a result. This is an area where ability to control partitioning of the graph over the cluster becomes critical.

Socrates supports three models of parallel processing over the distributed graph: DGraph, JGraph, and Neighborhood. Each type provides a different tradeoff between ease of algorithm implementation, parallelism of client code, and network communication.

*DGraph:* In the first model, Socrates provides clients with access to the DGraph class, which implements the Blueprints API and abstracts away the distributed nature of the underlying graph. Methods of the DGraph class are implemented with parallel calls to the underlying database where possible, but all results are sent back to the client machine and no client code runs on the Socrates cluster. This model of parallelism is suitable for developing analytics that need a global view of the graph and cannot benefit from being parallelized across the entire cluster (examples?).

*JGraph:* In the second model, Socrates allows clients to create processing jobs that can be submitted to the cluster to run in parallel on each node; each job is given access to the JGraph local to the node it is being run on (See Fig. X). A JGraph is another implementation of the Blueprints API that represents the partial graph stored on its local machine. Vertex iterators used in parallel jobs on JGraphs iterate over vertices that are local in that machine. However any questions asked about local vertices, such as getNeighbors() operation, retrieves all matching results independent of where they are located. Therefore the implementation of parallel jobs is quite similar to a regular standalone program. User has a clear view of the boundaries of the local graph and can limit operations on the local graph based on this information. This model of parallelism is suitable for developing analytics that can make use of a wide view of the graph as well as benefit from parallelism, such as sub-graph isomorphism. It is also useful if the graph can be partitioned into disjoint sub-graphs that are small enough to fit on one machine; in this case, it is trivial to use Socrates' locality control features to make sure the entire sub-graphs are placed on the same machine, allowing any algorithm that runs on a DGraph in the previous model to be parallelized.

*Neighborhood:* The final model of parallelism supported by Socrates is intended for algorithms that perform local computation only, such as centrality measurements or PageRank. Socrates provides an interface that allows clients to define a function that will be run in batch on every vertex in the graph. When the function is called, its input is a TinkerGraph (an in-memory implementation of Blueprints) that contains one vertex labeled "root", and may contain other elements that the client specifies when the function processing job is submitted. The client is able to specify whether the TinkerGraph should contain the root vertex's immediate neighbors (or in/out neighbors only in the case of a directed graph) and their edges with the root, as well as any properties of vertices or edges that should be fetched. The client's function is then able to write out new property values for the root node or any of its neighboring edges (future implementations will allow new neighboring vertices and edges to be added as well).

This model of parallelism is intended to make it easy to write local-computation graph analytics that take full advantage of the computing power of a cluster. Under the hood, Socrates takes care of running the client function in parallel on each node using as many threads as the hardware on that node will support, optimizing SQL queries and inserts, caching frequently-used values, and minimizing network communication between nodes.

Communication for parallel processing is provided by Java Messaging Service (JMS) using publish-subscribe method. In order to eliminate centralized communication and potential bottlenecks, each machine operates its own message broker. Every machine on the cluster is therefore on an equal footing. Parallelizing message handling also eliminates a potential single point of failure in the system. Jobs are executed in parallel on each machine and the results are, optionally, returned back to the client submitting the job request.

## IV. PERFORMANCE

In this section, we provide performance benchmarks demonstrating the scalability of Socrates in terms of ingest and parallelized analytics.

### A. Cluster configuration

Our cluster consists of 16 servers which are equipped with quad-core Intel Xeon E5-2609 2.5GHz processors, 64 GB 1600MHz DDR3 RAM, and two 4.0TB Seagate Constellation HDDs in RAID 0. The servers are running CentOS, and Socrates is using MySQL 5.5 with the TokuDB storage engine as its data store.

### B. Ingest

To measure our ingest speeds, we insert large randomly-generated Erdos-Renyi (E-R) graphs [http://www.citeulike.org/group/3072/article/1666220] into the Socrates cluster from an external machine. The E-R graphs we use here consist of 100-vertex connected components with an average of 1000 edges each, however the ingest speed of these graphs depends only on the number of vertices and edges, and not on the underlying structure of the graph.

We insert E-R graphs with a total size varying from 100,000 vertices and one million edges to 100 million vertices and one billion edges. In order to see how well our ingest speed scales as the size of a cluster grows, we have repeated the ingest benchmarks using only 2, 4, and 8 nodes in addition to the full 16.

Figure 5 shows the ingest speeds, expressed as elements inserted per second, for graphs and clusters of varying sizes.

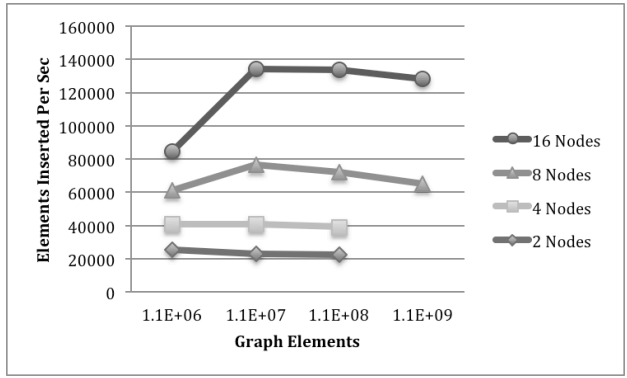

Fig. 5: Socrates insertion speeds for a randomly-generated Erdos-Renyi graph with an average of 10 edges per vertex, with varying cluster sizes.

We can see that, although the 8- and 16-node cluster does not have time to reach its top speed when ingesting the smallest graph tested (which takes them 18 and 13 seconds, espectively), ingest speeds hold steady even as the input graphs grow to over a billion elements. Part of the reason for this is our use of TokuDB[1] as the storage engine for MySQL. Previous versions of Socrates used the standard InnoDB engine, and suffered from severe I/O bottlenecks when the graph grew above roughly 200 million edges, likely due to InnoDB no longer being able to fit the interior nodes of the B-Tree in its in-memory buffer pool. TokuDB uses cache-oblivious lookahead arrays [http://dl.acm.org/citation.cfm?id=1248393] in place of B-Trees, and seems to avoid this problem.

Figure 6 shows the ingest speeds on a logarithmic scale for varying cluster sizes ingesting an E-R graph with 10 million vertices and 100 million edges.

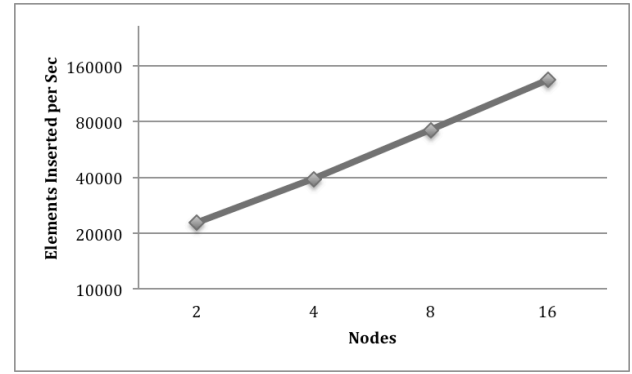

Fig. 6: Socrates insertion speeds (on a logarithmic scale) for an E-R graph with 10 million vertices and 100 million edges, with varying cluster sizes. On our cluster, Socrates has exhibited approximately linear ingestion speed-up as nodes are added.

We can see that Socrates ingest speed scales linearly to at least 16 nodes.

### C. Parallel Processing

To measure our parallel processing capability, we use a naive connected component algorithm implemented in the Neighborhood parallelism model described above. On its initial iteration, the algorithm assigns each vertex a *component* attribute equal to the smallest vertex id among itself and its neighbors. On subsequent iterations, the algorithm examines the *component* attribute of itself and its neighbors, and updates its *component* to be the smallest value in the examined set. The algorithm terminates when no vertex's *component* changes in an iteration.

Not that this benchmark is not necessarily the fastest method for computing connected components, however it is useful as a benchmark because Socrates must fetch each vertex's in and out neighbors along with property information for each node. Thus, the speed at which an iteration of this algorithm runs can give us an idea of the batch processing capability of the Neighborhood parallelism model.

We have run this algorithm on the same graphs that were ingested in the previous experiment, i.e., Erdos-Renyi graphs consisting of connected components with 100 vertices and an average of 1000 edges each, varying in total size from 1.1 million to 1.1 billion elements. Once again, we have repeated the experiments using only 2, 4, and 8 nodes in our cluster in addition to the full 16.

Figures 7 and 8 present our results of these experiments. The number of vertices processed per second, averaged over all iterations of the algorithm after the initial iteration, is on the y-axis. Note that processing a single vertex involves fetching its immediate neighborhood (an average

---
[1] http://www.tokutek.com/products/tokudb-for-mysql/

of 10 edges and vertices), as well as the *component* property for each vertex in the neighborhood.

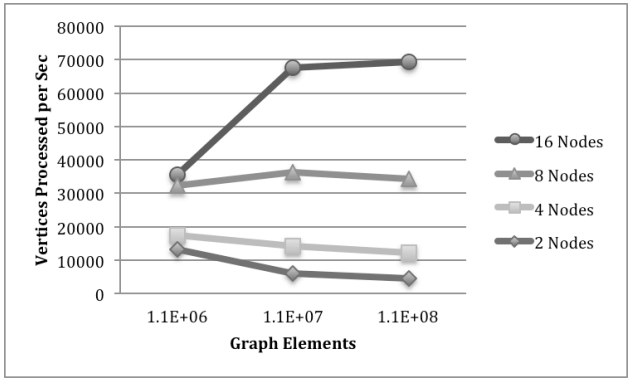

Fig. 7: Average processing speeds for an iteration of the connected component algorithm on an ER graph with an average of 10 edges per vertex.

These results are qualitatively similar to the ingest results. Figure 7 demonstrates that, once the graph is large enough to allow the 8- and 16-node clusters to reach full speed, processing speed holds steady up to graphs of over a billion elements.

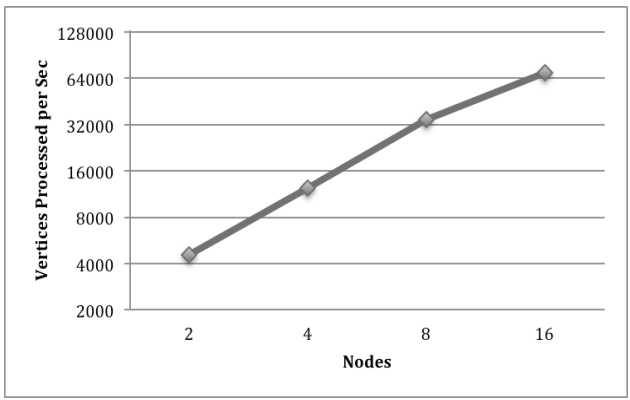

Fig. 8: Average processing speeds (on a logarithmic scale) for an E-R graph with 10 million vertices and 100 million edges, with varying cluster sizes. Again, we see approximately linear processing speed up as nodes are added.

Figure 8 demonstrates that processing speeds also scale in an approximately linear fashion up to a cluster of 16 nodes.

ACKNOWLEDGMENT

Visualizations in the paper were produced using Pointillist, a graph visualization software developed by J. Cohen. We acknowledge D. Silberberg and L. DiStefano for providing feedback and support during the development of Socrates.

REFERENCES

[1] P. Gupta, A. Goel, J. Lin, A. Sharma, D. Wang, and R. Zadeh, "WTF: The Who to Follow Service at Twitter," in *Proceedings of the 22Nd International Conference on World Wide Web*, Republic and Canton of Geneva, Switzerland, 2013, pp. 505–514.

[2] S. Salihoglu and J. Widom, "GPS: A Graph Processing System," in *Proceedings of the 25th International Conference on Scientific and Statistical Database Management*, New York, NY, USA, 2013, pp. 22:1–22:12.

[3] G. Malewicz, M. H. Austern, A. J. . Bik, J. C. Dehnert, I. Horn, N. Leiser, and G. Czajkowski, "Pregel: A System for Large-scale Graph Processing," in *Proceedings of the 2010 ACM SIGMOD International Conference on Management of Data*, New York, NY, USA, 2010, pp. 135–146.

[4] U. Kang, C. E. Tsourakakis, and C. Faloutsos, "PEGASUS: A Peta-Scale Graph Mining System Implementation and Observations," in *Ninth IEEE International Conference on Data Mining, 2009. ICDM '09*, 2009, pp. 229–238.

[5] P. Kalmegh and S. B. Navathe, "Graph Database Design Challenges Using HPC Platforms," in *High Performance Computing, Networking, Storage and Analysis (SCC), 2012 SC Companion:*, 2012, pp. 1306–1309.

[6] D. Dominguez-Sal, P. Urbón-Bayes, A. Giménez-Vañó, S. Gómez-Villamor, N. Martínez-Bazán, and J. L. Larriba-Pey, "Survey of Graph Database Performance on the HPC Scalable Graph Analysis Benchmark," in *Proceedings of the 2010 International Conference on Web-age Information Management*, Berlin, Heidelberg, 2010, pp. 37–48.

[7] J. P. Cedeño and K. S. Candan, "R2DF Framework for Ranked Path Queries over Weighted RDF Graphs," in *Proceedings of the International Conference on Web Intelligence, Mining and Semantics*, New York, NY, USA, 2011, pp. 40:1–40:12.

[8] R. Angles and C. Gutierrez, "Survey of Graph Database Models," *ACM Comput Surv*, vol. 40, no. 1, pp. 1:1–1:39, Feb. 2008.

[9] J. Hayes and C. Gutierrez, "Bipartite Graphs as Intermediate Model for RDF," in *The Semantic Web – ISWC 2004*, S. A. McIlraith, D. Plexousakis, and F. van Harmelen, Eds. Springer Berlin Heidelberg, 2004, pp. 47–61.

[10] S. Jouili and V. Vansteenberghe, "An Empirical Comparison of Graph Databases," in *2013 International Conference on Social Computing (SocialCom)*, 2013, pp. 708–715.

[11] N. Bronson, Z. Amsden, G. Cabrera III, P. Chakka, P. Dimov, H. Ding, J. Ferris, A. Giardullo, S. Kulkarni, H. Li, M. Marchukov, D. Petrov, L. Puzar, Y. J. Song, and V. Venkataramani, "TAO: Facebook's Distributed Data Store for the Social Graph," *USENIX Annu. Tech. Conf. ATC*, 2013.